# Visible Light-Driven C-C Coupling Reaction of Terminal Alkynes at Atmospheric Temperature and Pressure Reaction Conditions using Hybrid Cu$_2$O-Pd Nanostructures

*Samantha Stobbe, Ravi Teja A. Tirumala, Sundaram Bhardwaj Ramakrishnan, Marimuthu Andiappan\**

Affiliations:

School of Chemical Engineering, Oklahoma State University, Stillwater, Oklahoma 74078, USA.

*Corresponding authors, Email: <u>mari.andiappan@okstate.edu</u>

**ABSTRACT**

Carbon-carbon (C-C) coupling reactions are widely used reactions in the production of fine chemicals, agrochemicals, and pharmaceuticals. In our recent contribution (Green Chemistry, 2019, 21, 5284-5290), cuprous oxide ($Cu_2O$) nanospheres are shown to be efficient catalysts for C-C coupling reactions of terminal alkynes such as phenylacetylene. Specifically, $Cu_2O$ nanospheres are shown to successfully catalyze oxidative C-C homocoupling reaction of phenylacetylene to form a coupling product, diphenyl diacetylene (DPDA). However, these $Cu_2O$ nanocatalyst requires a relatively high temperature of ~110 $^0C$. Herein, we report that photocatalysts built on hybrid palladium nanoclusters decorated on $Cu_2O$ (i.e., $Cu_2O$-Pd) can utilize visible light as energy input and successfully catalyze oxidative homocoupling of phenylacetylene at atmospheric temperature and pressure reaction condition. This was displayed by the comparison of oxidative homocoupling reactions both in the presence and without the presence of visible light. Our findings indicate that photocatalysts through $Cu_2O$-Pd nanospheres present to be a logical substitute, moving toward more efficient and sustainable industrial processes.

**Introduction**

Development in photocatalysis has emerged in the last decade as an effective means for pollution mitigation, converting natural resources and pharmaceutical applications.[1–5] It has been key in removing pharmaceutical remnants from wastewater and the surrounding environment. Various reactions carried out in the industry use thermal energy at relatively high temperatures.[6,7] The present issue with thermal catalytic technologies is the reaction occurs in the ground state potential energy surface. It requires high operating temperatures and high pressures to overcome the activation barrier that occurs.[8] High amounts of heat being utilized can create numerous unwanted results and creates many safety issues. Visible light photocatalysts is being offered as a more efficient alternative than the use of thermal energy.[7] Through photocatalysts, industrial process could be performed using lower temperatures and gaining higher selectivity. $Cu_2O$ has been presented as a promising semiconductor because it is cost effective, earth abundant and it holds a small band gap of 2.0-2.2eV.[9–11] These band structures allow for the absorption of a broad part of the visible light spectrum to be absorbed. Plasmonic mental nanostructures exhibit very increased cross sections and are sensitive to the geometry and surrounding environment due to a property called localize surface plasmon resonance.[12–14] This property allows metal oxide semiconductors to exhibit enhanced photocatalytic activity as compared to semiconductor-only photocatalysts. A hybrid can prove to exhibit this property as well, but metals such as Ag and Au are expensive and not as earth abundant. This makes $Cu_2O$ a reasonable semiconductor for observing photocatalysis.[9,15]

**Experimental Details**

To synthesis the spherical Cu$_2$O nanoparticles, a microemulsion technique was used at room temperature. This created smaller nanoparticles sized with an average diameter of 30-50 nm. The detailed procedures for the synthesis and characterizations of Cu$_2$O nanoparticles are provided in our previous contribution (Green Chemistry, 2019, 21, 5284-5290). The Pd nanoclusters decorated on Cu$_2$O (Cu$_2$O-Pd) are prepared through chemical reduction synthesis method by using K$_2$PdCl$_4$ as the Pd precursor. The calibration curve was done before by hand from prepared reactant samples. Before the oxidative homocoupling reaction began, the nanoparticles were suspended in a mixture of water and dimethylformamide (DMF, 13.5mL, DI Water, 1.5mL). This was added to a 25mL flask, hooked up to a condenser and then added a thermocouple if needed. Air was constantly flowed throughout the reaction to maintaining the oxidative environment needed for oxidative C-C homocoupling (OHR) reactions. Potassium Carbonate (207mg) was added while stirring. To start the reaction, phenylacetylene (100uL) was added. The reaction was sampled regularly throughout the period to observe the conversion rates.[16–19]

**Results and Discussion**

Glaser-type oxidative homo-coupling reaction (OHR) of phenylacetylene (PA) was used to investigate the photocatalytic nature of the nano catalysts. The nanoparticles used in all reactions were prepared using microemulsion synthesis method.[16] A detailed description of both the microemulsion technique and the reaction procedure are provided in our previous contribution (Green Chemistry, 2019, 21, 5284-5290). Since Glaser-type oxidative homo-coupling reaction only has a single reactant and product, the conversion of the reaction was efficient to observe.

To be able to systematically compare the types of energy, oxidative homo-coupling reactions were carried out in the presence of $K_2CO_3$ with and without light. The solvent used was a mixture of 90% dimethyl fluoride (DMF) and 10% DI water.

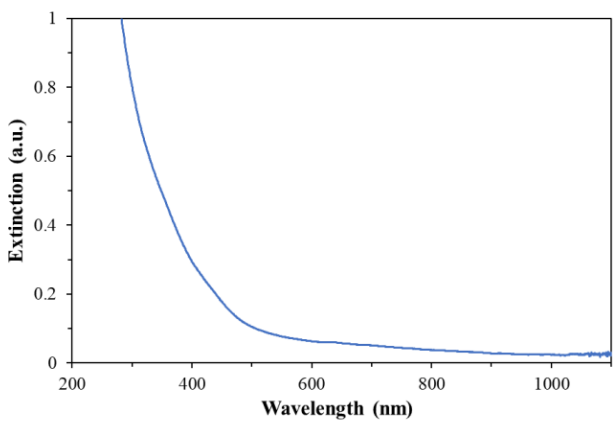 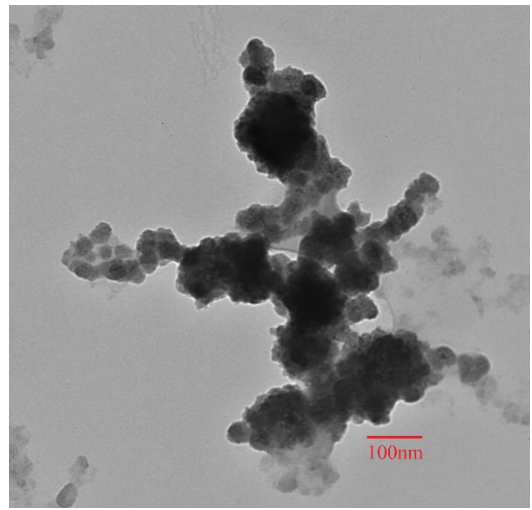

**Figure 1. (a)** UV-Vis extinction spectra of $Cu_2O$ nanospheres ranging from 30 nm-50 nm diameters $Cu_2O$ spherical nano catalysts were synthesized using the microemulsion technique. **(b)** TEM imaging showing nanospheres sizing at an average of 35nm.

The nanoparticles were then characterized through the UV-Vis and displayed by Figure 1a. To further characterize the nano catalysts, a sample was put through TEM imaging shown in Figure 2 (a-b) to confirm the sizing and shape of the nanospheres. Nanoparticles shown in Figure 1b averagely have a circular shape and rounded edges. Therefore, it can be confirmed that nanospheres around ~35 nm was being synthesized.

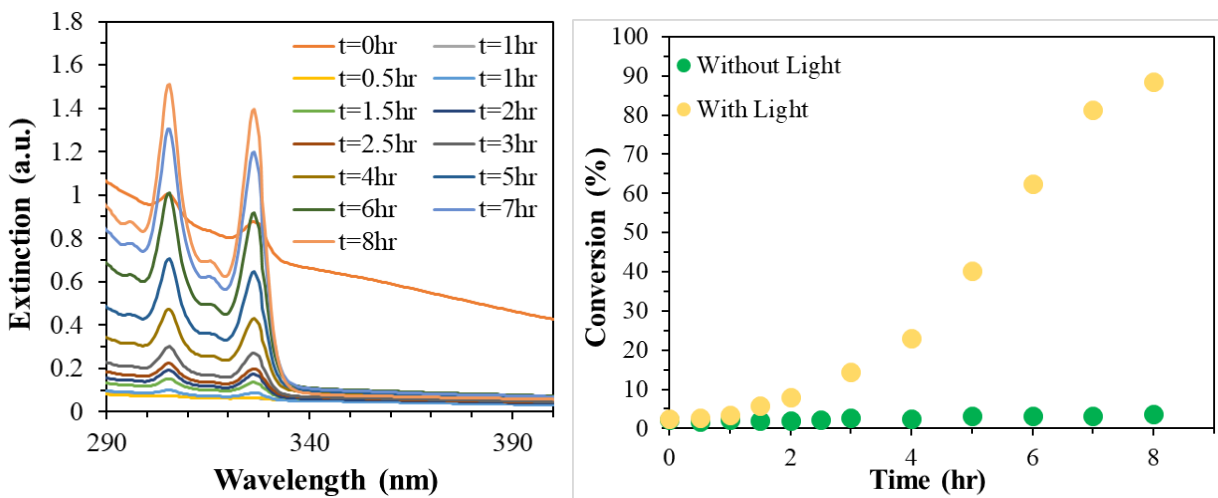

**Figure 2. (a)** UV-Vis extinction spectra of the reaction mixture after the addition of PA at different samples of time. **(b)** Reaction conversion as a function of reaction time for oxidative homocoupling on $Cu_2O$-Pd nanocatalysts with and without the effect of Light at the primary peak (326 nm) of DPDA.

The DPDA product was observed using characterization through running timed UV-Vis samples. Figure 3a shows primary and secondary extinction peaks for DPDA appear at ~326nm and ~305nm respectively. As the reaction continues, the amount of DPDA available inside of the reaction mixture increases with time. The OHR reaction over a period of 8 hours was complete with and without the presence of light. The extinction points of the primary peak (326nm) were used to quantify the conversion and then plotted in a comparison to the time as shown in Figure 3b. Observing Figure 3b, it can be noted that with the reaction is exposed to white light, the ending conversion show to be much higher than its respective counterpart. It can be concluded that the presence of light magnifies the reaction rate of the reaction when compared to without light.

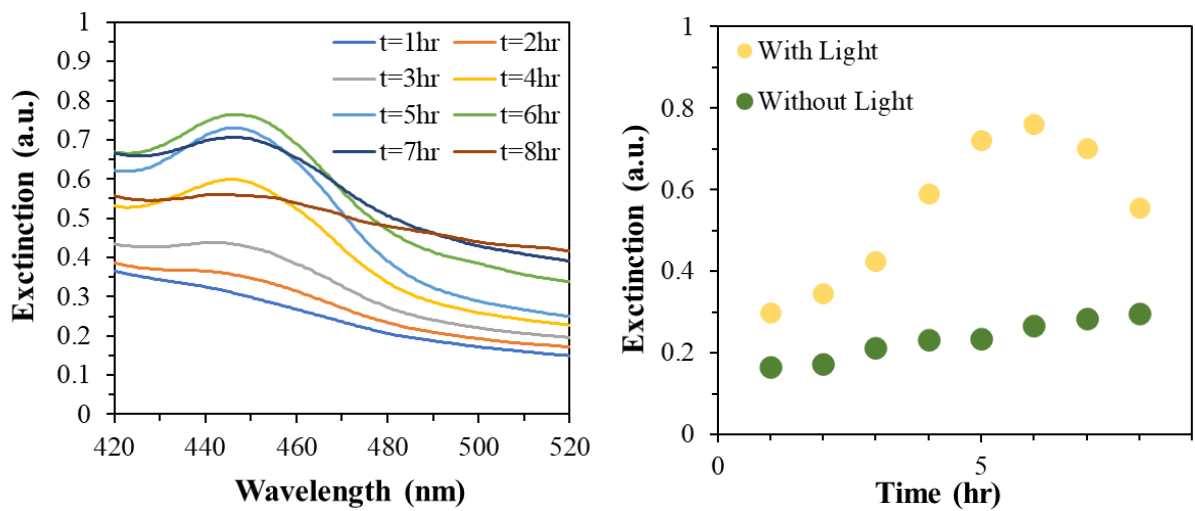

**Figure 3. (a)** UV-Vis extinction spectra of the reaction mixture after the addition of PA at different samples of time. **(b)** UV-Vis extinction spectra of the reaction mixture at 450nm as a function of time.

During both types of reactions, an extinction peak began to appear at ~450mn. A dilution of 100uL of the mixture of the reaction to 3mL of Ethanol was taking to observe the extinction peak and is displayed in Figure 4a more efficiently. These peaks indicate the formation of homogeneous Cu complexes, the intermediate species created in the reaction mixture between the reactant, PA, and the product, DPDA, while the reaction is taking place. It is expected and shown that there will be no intermediate species at the beginning of the reaction. As the reaction persist, the complex amount should increase and then decrease until there is no complex since the concentration of PA at the end of the reaction is relatively low. PA is a reactant to create the complex and therefore can be attributed to the expected decreasing trends. Figure 4b shows the comparison of the trend of complexes in both types of reactions as a function of time. Without light, the formation of Cu complex stays relatively linear over the 8-hour period. Because a great increase or the start of a decrease in extinction is never seen, it can be concluded that the rate of the reaction must be extremely slow. With light, the extinction shows the expect result of increasing and then decreasing in extinction. Compared to without light, the rate of reaction must be significantly higher.

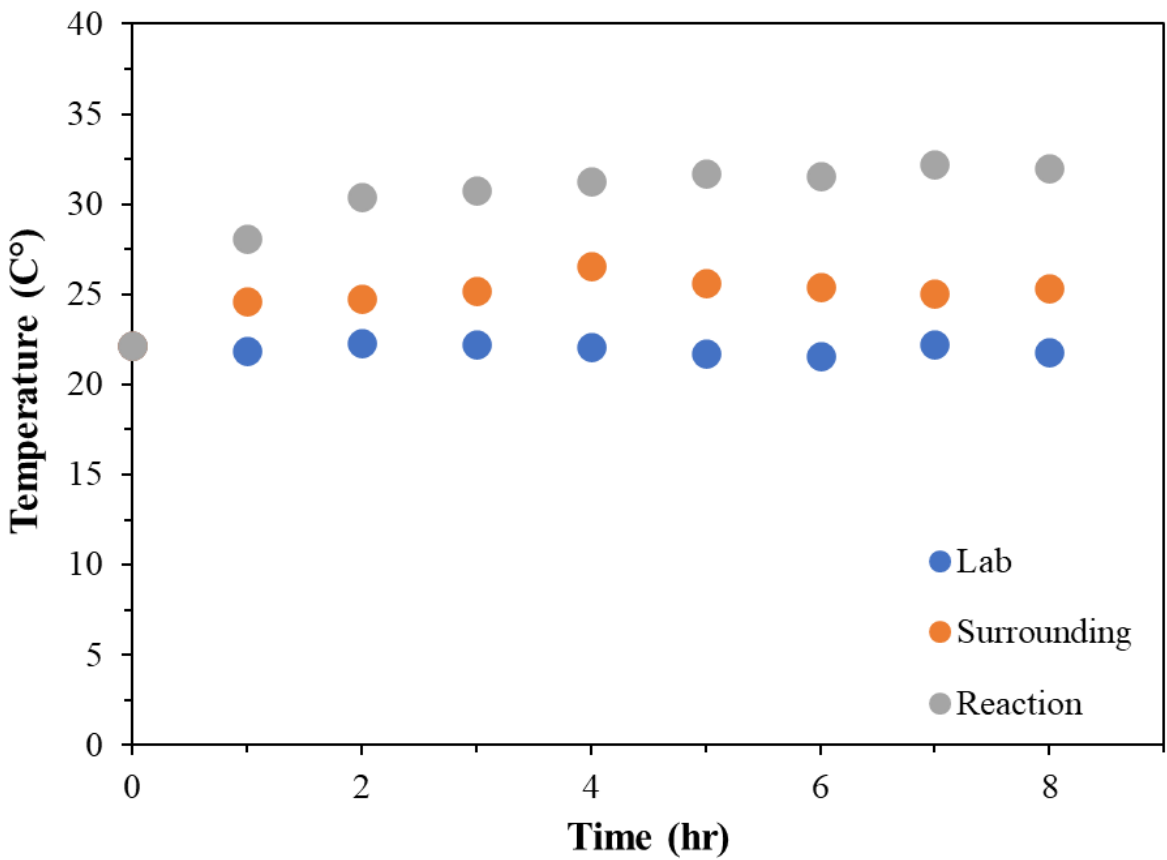

**Figure 4.** Measured temperature of the lab, surrounding the reaction and the reaction itself in of the reaction with the effect of light.

Thermal energy is another variable that is highly used inside of the industry for many energy and power reactions. In order to create a comparison, the same reaction was done at around ~35°C to observe the reaction rate just using thermal energy. The temperature was recorded using a handheld electronic thermometer throughout each reaction using light and was shown in Figure 5.

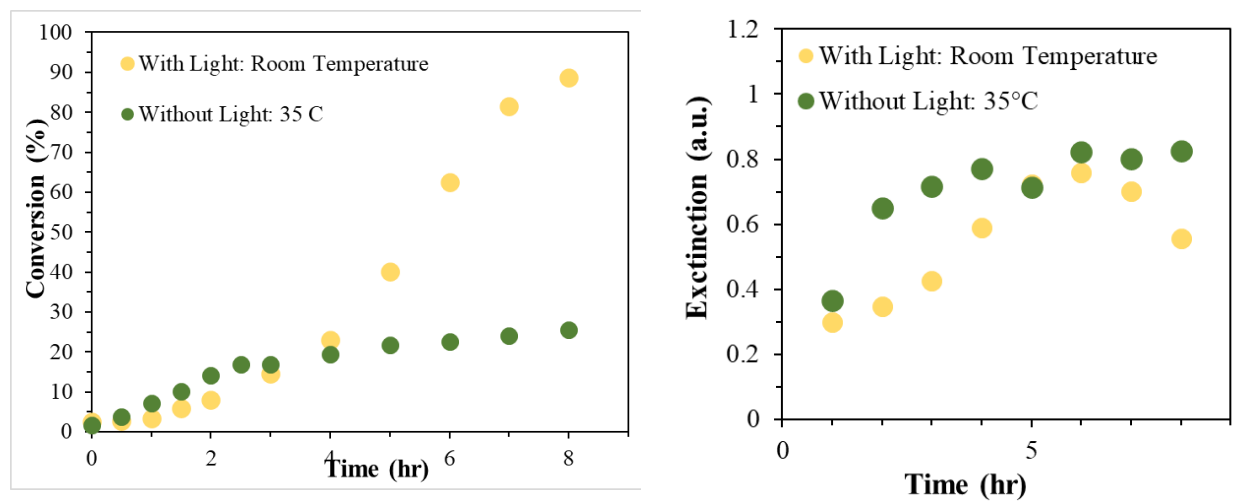

**Figure 5. (a)** Reaction conversion as a function of reaction time for oxidative homocoupling of PA on $Cu_2O$-Pd nanocatalysts without light heated to 35° C, in comparison to with light in which maximum temperature of ~35 °C is reached. **(b)** UV-Vis extinction spectra of the reaction mixture at 450nm as a function of time.

The rate of PA conversion is high under light on conditions in comparison to dark conditions performed at 35 °C (Figure 5a). we attribute the higher rate observed under light ON conditions to the combination of light induced heat effect and excited electron-driven photochemistry. This excited electron-driven photochemistry can occur via Menzel–Gomer–Redhead (MGR) mechanism listed in our recent review article.[20]

A reaction was done without light while being heated to 35 °C to simulate the reaction. As shown by Figure 6a, the reaction when heated shows similar conversion to that with of a reaction with light. Observing the trendline, the conversion rate of the reaction without light seems to be deactivating or decreasing in slope as the time increases. Comparing that to the reaction with light, the trendline is still increasing immensely and hypothetically will continue to hold that rate after the 8-hour period. Even though this may show similar conversions, after 8 hours the conversion of the reaction with light would show much greater conversion that the reaction being heated. Figure 4b confirms that theory being presented. The reaction without light shows a very linear complex formation compared to the reaction with light increases and decreases with time. As discussed before, this is an indicator of a slow reaction compared to a much faster reaction, respectively.

**Conclusions**

The results obtained by the UV-Vis extinction spectroscopy for the $Cu_2O$-Pd oxidative homocoupling reaction with and without the presence of light are consistent with the hypothesis presented. The higher calculated conversion rates as well as the consistent Cu complex trends furthers this statement. The rate of PA conversion is high under light on conditions. We attribute the higher rate observed under light ON conditions to the combination of light induced heat effect and excited electron-driven photochemistry. It can be stated that using hybrid catalyst built on metal oxide semiconductors for photocatalytic technologies would show to be effective against thermal energy technologies. The work and results presented in our study opens the door to more development for solar and pharmaceutical applications.

AUTHOR INFORMATION

*Corresponding authors, Email: mari.andiappan@okstate.edu